# Modulation of Electroosmotic Flow through Short Nanopores by Charged Exterior Surfaces

Chao Zhang,[1] Xiaomei Zhang,[1] Hongwen Zhang,[2,3] Zekun Gong,[2,3] Xiuhua Ren,[1] Mengnan Guo,[1] and Yinghua Qiu[2,3]*

1. School of Mechanical and Electronic Engineering, Shandong Jianzhu University, Jinan, 250101, China
2. Key Laboratory of High Efficiency and Clean Mechanical Manufacture of the Ministry of Education, State Key Laboratory of Advanced Equipment and Technology for Metal Forming, School of Mechanical Engineering, Shandong University, Jinan, 250061, China
3. Shenzhen Research Institute of Shandong University, Shenzhen, 518000, China

*Corresponding author: yinghua.qiu@sdu.edu.cn

**ABSTRACT**

Electroosmotic flow (EOF) through nanoporous membranes has broad applications in micro- and nanofluidic systems, particularly in biomedical diagnostics and chemical analysis. The use of short nanopores enables high fluid flux, and the presence of exterior surface charges can further enhance ion flux through short nanopores. Here, systematic simulations are conducted to explore the modulation of EOF by exterior surface charges. Our results indicate that charged exterior surfaces can provide an additional effective pathway for fluid flow, significantly increasing both the EOF velocity and output pressure. By analyzing the dependence of EOF velocity on the area of the charged exterior surface, we derive the effective width ($L_{cs\_eff}$) of the charged ring region extending beyond the pore boundary. This parameter is quantitatively examined under various nanopore configurations and applied conditions. $L_{cs\_eff}$ is found to be proportional to the pore diameter, surface charge density, and applied voltage, while inversely proportional to pore length and salt concentration. These findings provide valuable insights into the modulation of EOF by exterior surface charges and offer theoretical guidance for optimizing the structural and functional properties of nanoporous membranes in practical applications of EOF.

**Nomenclature**

Table 1 Symbols List

| Symbol | Meaning |
|---|---|
| $\varepsilon$ | Dielectric constant (C V$^{-1}$ m$^{-1}$) |
| $\zeta$ | Zeta potential (V) |
| $\mu$ | Solution viscosity (mPa·s) |
| $E_p$ | Electric field strength (mV nm$^{-1}$) |
| $C_i$ | Concentration of species (mM) |
| $d$ | Pore diameter (nm) |
| $F$ | Faraday constant (C mol$^{-1}$) |
| $D_i$ | Diffusion coefficient of species $i$ (m$^2$ s$^{-1}$) |
| $I$ | Total current (nA) |
| $\nabla$ | Gradient (grade) |
| $p$ | Pressure (bar) |
| $\varphi$ | electrical potential (V) |
| $J_i$ | ionic flux of species $i$ |
| $k_B$ | Boltzmann constant (J K$^{-1}$) |
| $L$ | Pore length or charged surface length (nm) |
| $w$ | The axial flow component (m s$^{-1}$) |
| $R$ | Gas constant (J kg$^{-1}$ K$^{-1}$) |
| $T$ | Temperature (K) |
| $u$ | Velocity of solution |
| $z_i$ | Valence of species $i$ |
| $Q$ | Flow flux (m$^3$ s-1) |
| $A$ | Cross-sectional area of the nanopore (m$^2$) |
| *v* | Velocity |

<div align="center">Table 2 Subscript List</div>

| Symbol | Meaning |
|---|---|
| EOF | Electroosmosis Flow |
| ICS | The nanopore with only charged inner-pore surfaces |
| ACS | The nanopore with all charged surfaces |
| en | Entrance |
| ex | Exit |
| cs | Charged exterior surface |
| cs-eff | Effective charged exterior surface |
| i | Cation or anion |

I. Introduction

Electroosmotic pumps, based on various nanoporous membranes, have found widespread application in microfluidic analysis, micro/nano liquid chromatography, chemical analysis, drug delivery, and biomolecule separation.[1] As the functional units of nanoporous membranes, nanopores offer a versatile platform for the investigation of ion and fluid transport through porous membranes.[2] At the membrane-liquid interfaces, solid surfaces can acquire charges through mechanisms such as the dissociation of surface chemical groups.[3] For example, the surfaces of silica nanopores and polyethylene terephthalate nanopores can become negatively charged due to the deprotonation of silanol groups and carboxyl groups, respectively, when in contact with aqueous solutions.[4-6] These induced surface charges, which are pH-dependent,[7] play a critical role in modulating the ionic and fluidic behaviors inside nanopores, including electroosmotic flow,[8, 9] surface-charge-governed conductance,[10, 11] selective ion transport,[12, 13] and ionic current rectification.[14-17]

At the solid–liquid interface, electrostatic interactions between surface charges and ions lead to the accumulation of counterions near charged pore walls, i.e., resulting in the formation of electric double layers (EDLs).[3, 7, 18] When a voltage is applied across the nanopore, the tangential electric field motivates the migration of counterions inside EDLs, thereby inducing the electroosmotic flow (EOF) due to ionic hydration.[19-21] According to the classical description based on the Helmholtz-Smoluchowski equation,[3, 22, 23] the EOF velocity ($v_{EOF}$) through micropores can be predicted with $v_{EOF}=-\varepsilon\zeta E_p/\mu$, where $\varepsilon$, $\zeta$, $\mu$, and $E_p$ are the dielectric constant, zeta potential, solution viscosity, and electric field strength, respectively. Both nanofluidic experiments and simulations, the modulation of EOF through nanopores by parameters of nanopores and applied conditions have extensively explored how EOF through nanopores is modulated by nanopore parameters and applied conditions.[9, 22, 24-26] From classical theoretical predictions and previous studies, it is evident that the EOF velocity is closely dependent on the pore diameter,[5, 27] membrane thickness,[22, 28, 29] surface charge properties,[9, 30, 31] applied voltage,[9, 25, 27] and solution concentration.[5, 27, 32, 33]

As a key electrokinetic phenomenon, EOF through nanopores has garnered significant attention. Recent studies have revealed a variety of intriguing EOF behaviors. For instance, using molecular dynamics simulations, Rezaei et al.[34] examined EOF in silicon nanochannels immersed in NaCl solutions. They found that as the surface charge density increases, the EOF velocity initially increases, then decreases, and eventually reverses direction due to the appearance of charge inversion.[19, 35] With conical glass nanopipettes, Laohakunakorn et al.[36] investigated the fluid flow outside the glass nanopore using optical tweezers. Their results indicated that at concentrations below 1 mM, the EOF reversed direction. Additionally, under a viscosity gradient across SiN nanopores, a reversed EOF was observed, attributed to the slower depletion of co-ions compared to counterions along the nanopore.[26] Using porous membranes with conical nanopores, Wu et al.[37] investigated the EOF rectification under alternating voltages, which enabled the development of their alternating current electroosmotic pump. Based on the EOF-controlled solution distribution inside micro/nanopores,[38] ionic diodes with controllable ionic current rectification ratios had been constructed,[39] which hold potential applications in measuring solution viscosity[39] and surface charge density of pore walls.[5]

Inside nanopores, based on the Hagen-Poiseuille equation, the permeability of liquids is inversely proportional to the pore length,[29] which means short nanopores are advantageous for achieving high liquid flux.[31] For example, leveraging an ultrathin membrane with a thickness of approximately 75 nm, Yang et al.[25] developed a high-efficiency electroosmotic pump that achieved a normalized EOF rate of 172.90 mL/min/cm/V at 1 V, demonstrating significant potential for drug delivery and microchip sample injection.[24] With nanofluidic experiments, Nilsson et al.[40] successfully explored the modulation of ionic current through silicon nitride nanopores by exterior surface charges achieved by controlled oxide deposition. Later, simulation studies have shown that for nanopores with a length below ~200 nm, exterior surface charges play an important role in modulating ion transport.[41-45] The EDLs formed near charged exterior surfaces can provide additional pathways for counterion transport, significantly enhancing the flow of counterions under electric fields[16, 41, 44] or concentration gradients.[42, 46] In these cases, a substantial number of counterions from the reservoir are attracted to the EDLs near exterior charged surfaces on the entrance side and migrate into the nanopore along these exterior surfaces. On the exit side, charged exterior surfaces increase the area for ionic diffusion from the pore opening to the bulk solution.[41, 46]

Given that exterior surface charges significantly enhance ionic flux through short nanopores, EOF is expected to exhibit a strong dependence on these charges. In this work, we investigate the modulation of EOF through short nanopores by charged exterior surfaces. Systematic simulations have been performed to examine EOF under varying nanopore parameters and external conditions. By analyzing the electric field strength and ion concentration inside the pore, we identify the regulation mechanism of EOF by charged exterior surfaces. The effective width of the charged exterior region around the nanopore, i.e., the minimum width of the charged region for the nanopore to reach its best performance, is determined by adjusting the charged exterior area near the pore orifice. Additionally, the quantitative dependence of the effective charged width is explored on nanopore parameters and external conditions.

Note that in this work, we focused on the detailed characteristics of EOF through nanopores. Though charged nanopores present strong ionic selectivity,[12, 13] the exploration of ionic selectivity is beyond the scope of this work, which can be conducted in the following work. Due to the symmetric shape and charge distribution of cylindrical nanopores, ion current rectification [16, 47] is absent under applied electric fields.

**II. Simulation Details**

Simulation models were constructed using COMSOL Multiphysics to investigate the ion and fluid transport inside nanopores under applied voltages.[41, 48, 49] Coupled Poisson-Nernst-Planck (PNP) and Navier-Stokes (NS) equations were solved to account for ion distribution near charged pore walls, as well as the transport of ions and fluid in aqueous solutions, as described by Eqs. 1-4.[50-55]

$$\varepsilon \nabla^2 \varphi = -\sum_{i=1}^{N} z_i F C_i \quad (1)$$

$$\nabla \cdot \boldsymbol{J}_i = \nabla \cdot \left( C_i \boldsymbol{u} - D_i \nabla C_i - \frac{F z_i C_i D_i}{RT} \nabla \varphi \right) = 0 \quad (2)$$

$$\mu \nabla^2 \boldsymbol{u} - \nabla p - \sum_{i=1}^{N} (z_i F C_i) \nabla \varphi = 0 \quad (3)$$

$$\nabla \cdot \boldsymbol{u} = 0 \quad (4)$$

where $\varepsilon$ and $\mu$ represent the dielectric constant and viscosity of solutions, respectively. $\nabla$, $\varphi$, $N$, $F$, and $R$ denote the gradient operator, electrical potential, number of ionic species, Faraday constant, and gas constant, respectively. $\boldsymbol{u}$, $p$, and $T$ refer to the fluid velocity, pressure, and temperature. $z_i$, $C_i$, $\boldsymbol{J}_i$, and $D_i$ are the valence, concentration, ionic flux, and diffusion coefficient of ionic species $i$ (including cations and anions), respectively.

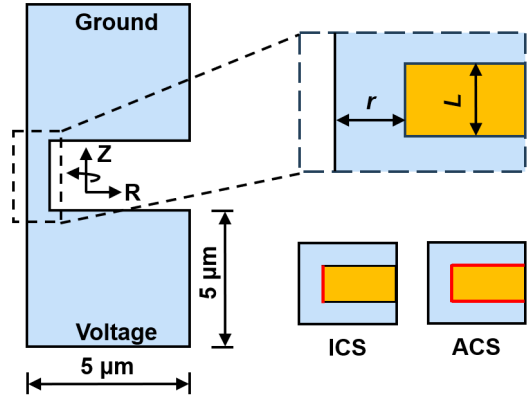

Figure 1. Schematic diagrams of the simulation models. Both the radius and height of the reservoir are 5 μm. The zoomed-in part shows the nanopore, with its length and radius denoted as *L* and *r*, respectively. Two simulation models with different surface charge configurations are considered, i.e., the nanopore with charged inner-pore surfaces (ICS) and the nanopore with all charged surfaces (ACS). Charged surfaces are shown in red.

As illustrated in Figure 1, a nanopore is located between two reservoirs, each having a radius and height of 5 μm. Note that the reservoir should be larger than 800 nm to avoid the negative impact of its size on the results.[56] To study the effect of the charged exterior surface on the ion and fluid transport in the nanopore, two simulation models were established with different surface charge configurations, i.e., the nanopore with only charged inner-pore surfaces (ICS) and the nanopore with all charged surfaces (ACS).[41, 46] With both simulation models, EOF characteristics were examined systematically under the influence of various nanopore parameters, i.e., the pore length, pore diameter, and surface charge density. The effects of external conditions, such as salt concentration and applied voltage, on EOF were also considered. In our simulations, the nanopore diameter and length were varied between 3 and 50 nm, and between 10 to 200 nm, respectively.[41, 42, 46] The default pore diameter and length were 10 nm and 50 nm.[25, 26] The nanofluidic system was filled with a KCl solution of a concentration ranging from 1 to 100 mM, with 100 mM as the default value. The diffusion coefficients of $K^+$ and $Cl^-$ were set to $1.96 \times 10^{-9}$ and $2.03 \times 10^{-9}$ m$^2$/s, respectively.[57] For charged pore walls, the surface charge density was changed from $-0.04$ C/m$^2$ to $-0.16$ C/m$^2$.[3, 58, 59] The applied voltage across the nanopore ranged from 0.05 to 2 V. The default surface charge density and voltage were set to $-0.08$ C/m$^2$ and 1 V, respectively. The system temperature was maintained at 298 K, and the relative permittivity of water was set to 80. Detailed boundary conditions are listed in Table S1. Please note that the choice of electrostatic and hydrodynamic boundary conditions was based on the properties of membrane materials, such as silicon, silicon nitride, and 2D materials,[60] which are usually considered for the fabrication of thin nanopores. For those thin nanopores immersed in aqueous solutions, both inner and exterior walls share the same properties as the bulk materials.

The flow flux ($Q$) was calculated by integrating the axial flow component ($w$) within the nanopore over the reservoir boundary, as described by Eq. 5.[41, 50, 51, 61]

$$Q = \int_s w ds \qquad (5)$$

where $S$ represents the reservoir boundary.

The same mesh strategy was employed as used in our previous simulations (Figure S1).[13, 41, 49] Given that EDLs significantly influence ion transport and, consequently, the EOF inside nanopores, the mesh size was set to 0.1 nm on the inner-pore walls and exterior surfaces within 3 μm beyond the pore boundary. For the remaining exterior surfaces, a larger mesh size of 0.5 nm was used to reduce computational cost. A mesh independence test was conducted as shown in Figure S1. With our mesh setting, the detailed ion and fluid transport can be investigated inside the EDL regions. Note that for the cases with charged exterior surfaces longer than 3 μm, a 0.1 nm mesh was used on the charged exterior surfaces, and a 0.5 nm mesh was used for the remaining exterior surfaces.

## III. Results and discussion

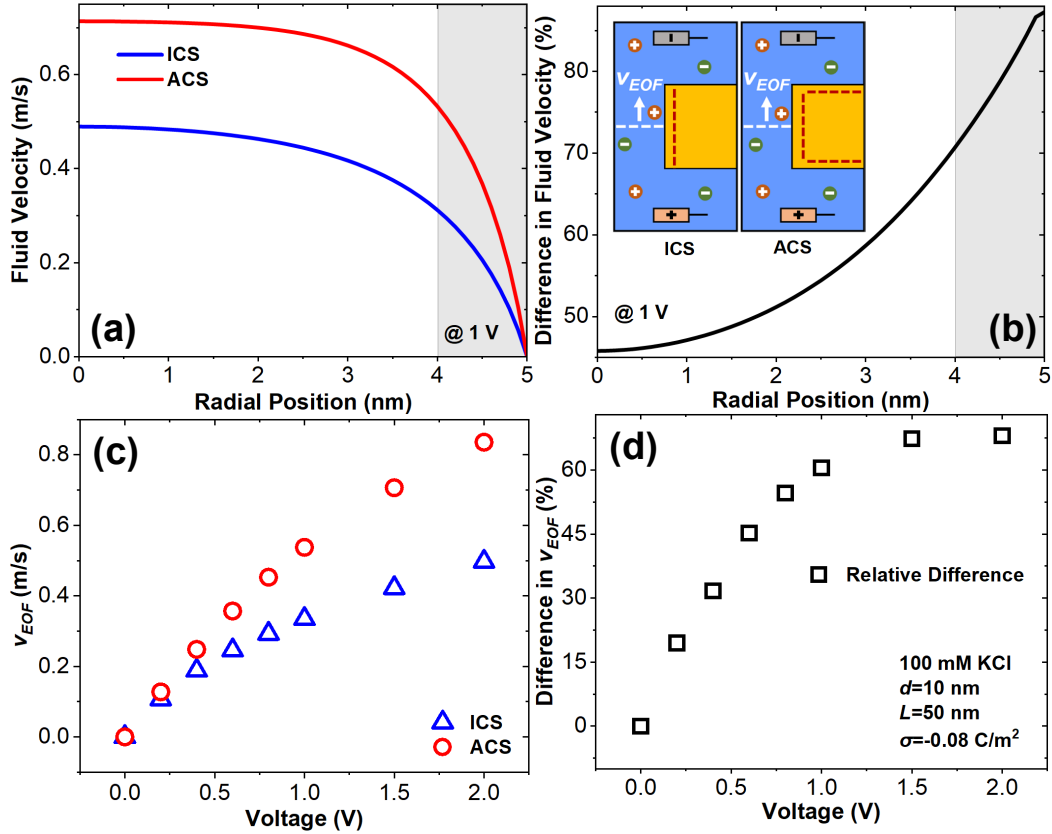

Figure 2. Effects of exterior surface charges on the electroosmotic flow (EOF) velocity inside nanopores under varying applied voltages. (a) Radial distribution of the EOF velocity at the central cross-section of the nanopore for the ICS and ACS cases. ICS and ACS cases denote the nanopore with only charged inner-pore surfaces and that with all charged surfaces, respectively. (b) Relative difference in the radial EOF velocity between ICS and ACS cases at 1 V. The inset shows the location of the central cross-section of the nanopore, with the light grey region corresponding to the EDLs. (c) Average EOF velocity at the pore center cross-section for the ICS and ACS cases, calculated with Eq.

6. (d) Relative difference in the average EOF velocity in ICS ($v_{ICS}$) and ACS ($v_{ACS}$) cases under different applied voltages, calculated by $(v_{ACS}-v_{ICS})/v_{ICS} \times 100\%$. The pore length is 50 nm, and the pore diameter is 10 nm. The surface charge density is −0.08 C/m². The solution is 100 mM KCl.

Considering that the voltage provides the most convenient way for the adjustment of the flux of electroosmotic pumps,[9, 27] we first investigated the modulation of exterior surface charges on the EOF velocity ($v_{EOF}$) under varying applied voltages. In both the ICS and ACS cases, the radial fluid velocity profile and the average EOF velocity are evaluated at the central cross-section of the nanopore. Figure 2a plots the radial distributions of $v_{EOF}$ in the ICS and ACS cases. At the solid surface, the fluid velocity is 0 due to the non-slip condition. The velocity increases sharply inside the EDL regions, and gradually approaches saturation at ~2 nm away from the surface. In the center of the nanopore, the maximum of $v_{EOF}$ reaches ~0.71 m/s and ~0.49 m/s in the ACS and ICS cases, respectively. Figure 2b illustrates the relative difference in the radial distribution of $v_{EOF}$ at the central cross-section of the nanopore. Inside the EDL region, the EOF velocity in the ACS case is higher than that in the ICS case by at least ~70%. As the distance from the pore center decreases, the difference in EOF velocity decreases and saturates at ~46%. This behavior is attributed to the viscous effect of the fluids and the diminishing influence of surface charges further from the pore walls. These results suggest that exterior surface charges significantly enhance EOF velocity within the nanopore.

To quantify the effect of exterior surface charges on EOF, the average $v_{EOF}$ through the nanopore is calculated using Eq. 6.

$$v_{EOF} = \frac{Q}{A} \tag{6}$$

where $A$ is the cross-sectional area of the nanopore.

Figure 2c shows the average EOF velocity at various voltages for both the ICS and ACS cases. As the applied voltage increases from 0 to ~0.5 V, the average $v_{EOF}$ increases approximately linearly in both cases, which agrees well with the prediction of the Helmholtz-Smoluchowski equation.[3, 22] Because the directional ion transport inside EDLs motivates the fluid flow, a higher voltage induces a larger $v_{EOF}$. At 0.4 V, the velocities are ~0.19 m/s for the ICS case and ~0.25 m/s for the ACS case. As the voltage rises further to 2 V, the rate of increase in both average $v_{EOF}$ slows, reaching ~0.50 m/s and ~0.84 m/s at 2 V for the ICS and ACS cases, respectively. This trend agrees well with the appearance of limiting current under high voltages.[41] The lower concentration of counterions at the pore entrance induces slower enhancement of $v_{EOF}$ as the voltage rises(Figure S2). The variation in EOF velocity with applied voltage demonstrates that voltage can be directly adjusted to modulate EOF through nanopores.

The enhancement in EOF velocity due to external surface charges can be quantified by $(v_{ACS}-v_{ICS})/v_{ICS}\times100\%$, in which $v_{ACS}$ and $v_{ICS}$ are the average EOF velocity in the ACS and ICS cases, respectively. As shown in Figure 2d, as the voltage increases from 0 to ~1 V, the relative increase in EOF velocity grows almost linearly to ~60.5%. Beyond 1 V, the relative increase gradually saturates, reaching ~68% as the voltage rises from 1 to 2 V. This trend presents the voltage-dependent promotion in $v_{EOF}$ by external charged surfaces. This can result from the transport capability of the additional effective pathway for fluid flow (see below) enabled by 5-μm-wide external charged surfaces. At the applied voltage of less than ~1 V, exterior charged surfaces can enhance more fluid transport under higher voltages. However, when the voltage exceeds ~1 V, the transport capability gets saturated and the promotion in $v_{EOF}$ stays at its maximum value.

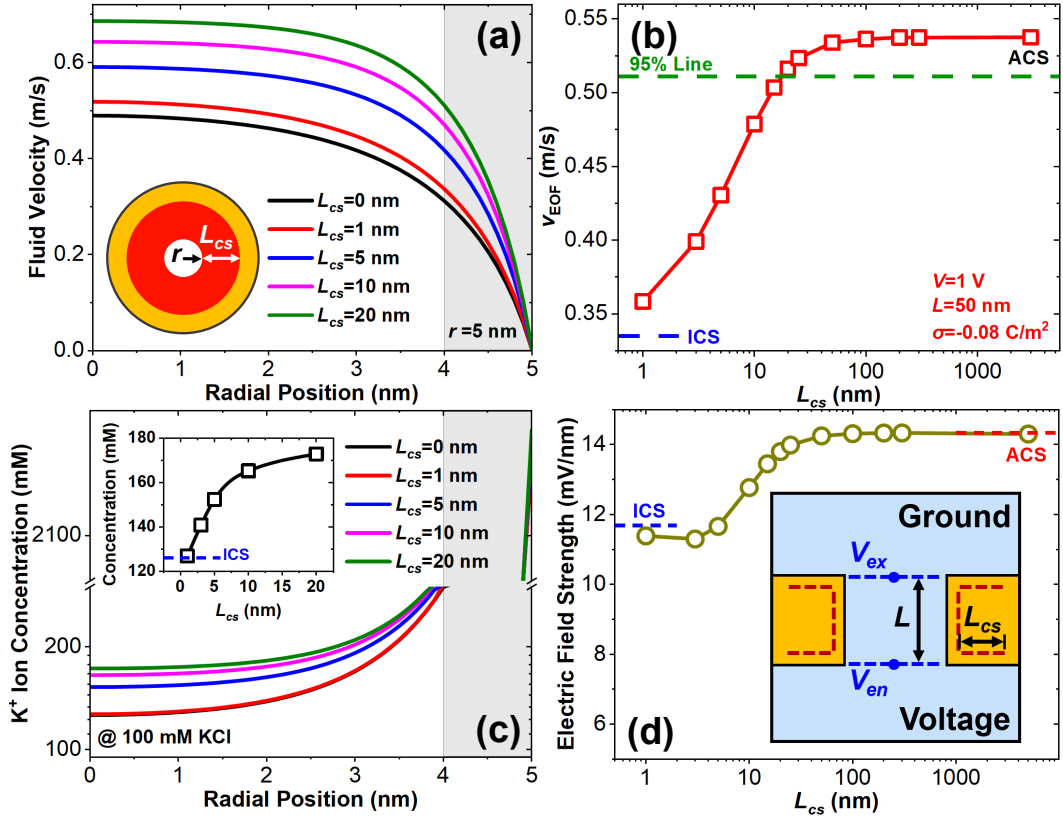

Figure 3. Characteristics of fluid and ion transport through a nanopore with various widths ($L_{cs}$) of the charged region. (a) Radial distributions of the electroosmotic flow (EOF) velocity at the central cross-section of the nanopore. (b) Average EOF velocity through the nanopore. The blue dashed line denotes the velocity under the ICS condition, and the green dashed line marks 95% of the velocity obtained in the ACS case. ICS and ACS cases denote the nanopore with only charged inner-pore surfaces and that with all charged surfaces, respectively. (c) Radial distributions of cation concentration at the central cross-section of the nanopore. The inset shows the concentration value of counterions at the pore axis. The blue dashed line corresponds to the value in the ICS case. (d) Electric field strength inside the nanopore. The inset shows the entrance and exit potentials, $V_{en}$ and $V_{ex}$. The electric field strength is calculated as $(V_{en}-V_{ex})/L$.

Statistical analysis of the average $v_{EOF}$ inside the nanopore reveals that the fluid flow in the ACS case is significantly higher than in the ICS case. To explore the mechanism underlying the flow enhancement caused by exterior surface charges, simulations were conducted with the nanopore containing charged ring regions at both ends. The width of the charged regions is denoted as $L_{cs}$, shown by the inset of Figure 3a. As $L_{cs}$ increases from 0 to 5 µm, the simulated nanopore transitions from the ICS case ($L_{cs}$=0 nm) to the ACS case ($L_{cs}$=5 µm).[41, 42, 46] Please note that on both sides of the membrane, the charged regions share the same value of $L_{cs}$.

Figure 3a illustrates the radial distributions of the EOF velocity in the central radial cross-section of the nanopore for various $L_{cs}$ values. As the $L_{cs}$ increases from 0 to 20 nm, the $v_{EOF}$ inside the nanopore is significantly enhanced. Specifically, the $v_{EOF}$ at the pore center can be increased from ~0.49 m/s to ~0.69 m/s, by ~41%. To quantitatively evaluate the modulation of $L_{cs}$ on the fluid flow, Figure 3b plots the average $v_{EOF}$ through the nanopore with $L_{cs}$ varying from 0 to 5 µm, which indicates that the average $v_{EOF}$ gradually increases from ~0.33 to ~0.51 m/s. Here, the effective width of the charged exterior region ($L_{cs\_eff}$) is defined as the value of $L_{cs}$ that delivers 95 % of the maximum flow rate achieved with $L_{cs}$ =5 µm. As shown in Figure 3b, for the 50-nm-long nanopore with 10 nm in diameter, $L_{cs\_eff}$ is ~20 nm. Note that the 95% threshold was chosen primarily because the negligible error range in experiments is generally within 5%. Of course, this standard can be adjusted as needed, such as to 97% or 90%.

Considering that EOF is induced by the directional movement of counterions, the ion concentration distribution deserves detailed investigation to uncover the underlying mechanism of the EOF enhancement by exterior surface charges.[3] Additionally, the distribution of counterions along the pore axis can influence the effective electric field strength.[26] Then, the ion concentration distribution at the center cross-section of the nanopore and the electric field strength across the nanopore are examined under varying $L_{cs}$ values. From Figure 3c, with $L_{cs}$ increasing from 0 to 20 nm, both the cation concentration inside the EDL region and at the pore center experience significant enhancement. At the center of the nanopore, the concentration of cations is ~126 mM in the ICS case, increasing to ~173 mM when $L_{cs}$ reaches 20 nm, which corresponds to a ~37% increase, as indicated in the inset.

Under applied electric fields, the selective transport of ions through the nanopore induced by surface charges can lead to ICP.[62] The formation of ICP can, in turn, modulate the effective electric field strength across the nanopore.[41] To assess this, we measured the potential values at the center of both the entrance and exit of the nanopore, denoted as $V_{en}$ and $V_{ex}$, respectively, as shown in the inset of Figure 3d. The electric field strength is calculated by the potential difference across the nanopore divided by the pore length, i.e., $(V_{en}-V_{ex})/L$. From Figure 3d, for $L_{cs}$ values less than ~3 nm, the electric field strength remains nearly constant. However, as $L_{cs}$ increases from 5 to ~20 nm, the electric field strength rises from ~11.7 to ~13.8 mV/nm, representing an increase of ~17.9%. When $L_{cs}$ exceeds ~25 nm, the electric field strength becomes relatively insensitive to further increases in $L_{cs}$. In the nanopore system, with $L_{cs}$ increasing, exterior charged surfaces provide an efficient transport pathway for counterions to migrate from the bulk to the nanopore entrance and from the pore exit to the bulk. The enhanced ion concentration at the pore entrance leads to a significant reduction in access resistance, which, in turn, enhances the electric field strength within the nanopore.[26, 41]

Based on the above statements, as $L_{cs}$ increases, both the enhanced cation concentration and the axial electric-field strength contribute to the increase in EOF within nanopores with exterior surface charges.

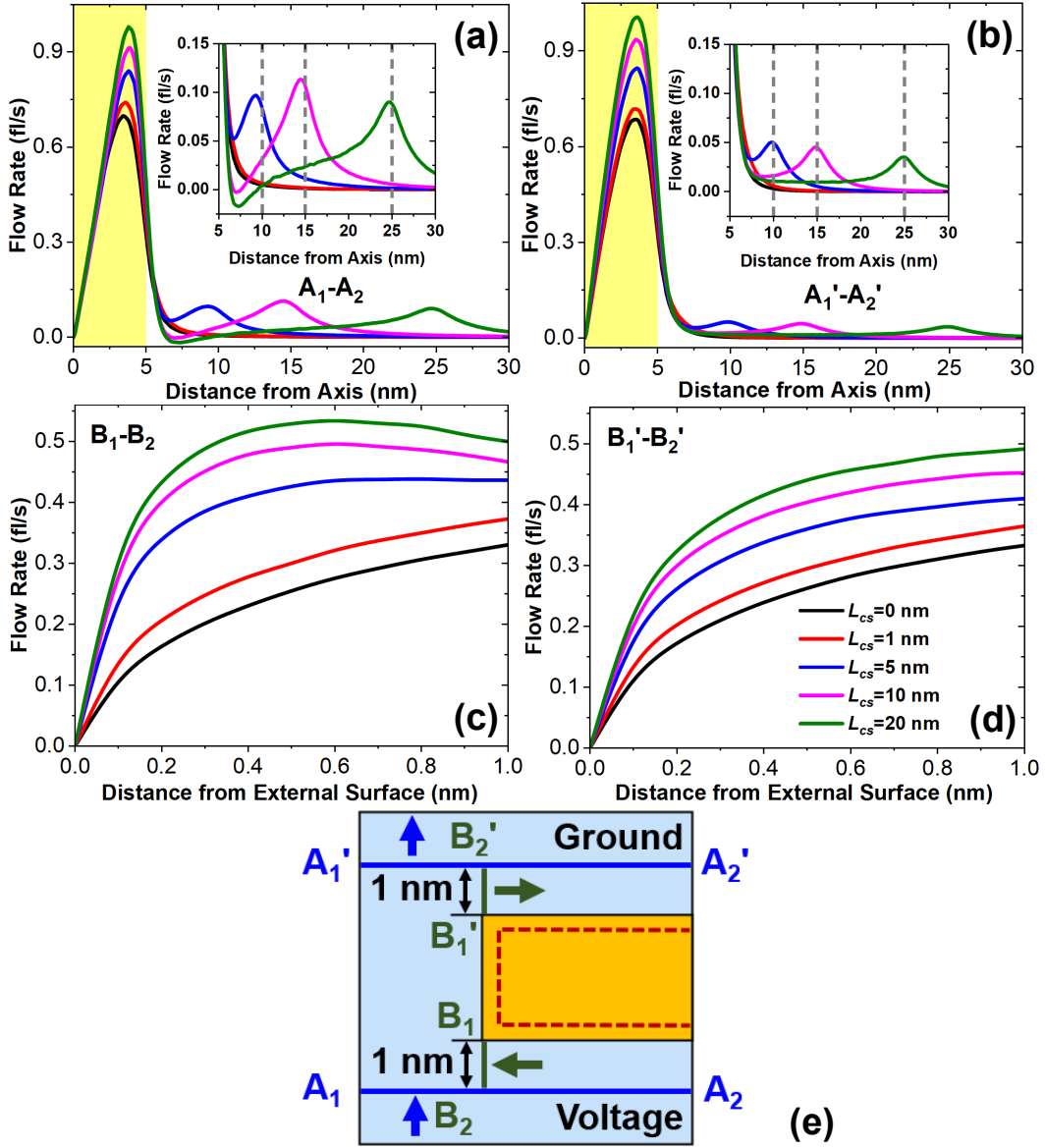

Figure 4. Flow rate distribution through the nanopore containing charged ring regions at both ends, with the width of the charged region denoted as $L_{cs}$. (a-b) Flow rate distributions in the horizontal planes $A_1$-$A_2$ (a) and $A_1'$-$A_2'$ (b). The light yellow part

represents the region above the nanopore opening. (c-d) Flow rate distributions in the vertical planes of $B_1$-$B_2$ (c) and $B_1'$-$B_2'$ (d). (e) Schematic diagram illustrating the planes where the flow rate distribution is obtained. Horizontal planes $A_1$-$A_2$ and $A_1'$-$A_2'$ are located 1 nm above the exterior membrane surface. The selection of 1 nm is due to that the Debye length is ~1 nm in 100 mM KCl solutions. $B_1$-$B_2$ and $B_1'$-$B_2'$ denote 1-nm-long vertical surfaces along the inner pore wall at both pore ends. Arrows indicate the direction of fluid flow.

To investigate how charged exterior surfaces enhance the fluid flow inside nanopores, we analyzed the distributions of the flow rate at different positions near both ends of the nanopore for varying $L_{cs}$ values (Figure 4).[41, 42] As shown in Figure 4e, planes $A_1$-$A_2$ and $A_1'$-$A_2'$ are positioned 1 nm above the exterior membrane surface. The selection of 1 nm is due to that the Debye length is ~1 nm in 100 mM KCl solutions. Considering that EOF is induced by the directional ion transport inside EDLs, the parallel flow rate is also considered, i.e., the fluid flow along exterior surfaces. Vertical surfaces $B_1$-$B_2$ and $B_1'$-$B_2'$ are located along the inner surface, extending 1 nm outward from the exterior surfaces of the nanopore. The local flow rate is obtained by integrating the axial or radial fluid velocity components over a 0.1 nm segment to investigate the detailed flow characteristics inside EDLs.

From Figures 4a and 4b, flow rate distributions along planes $A_1$-$A_2$ and $A_1'$-$A_2'$ illustrate the fluid flow along the pore axis. At both the pore entrance and exit, the primary flow rate is concentrated near the pore orifice within a radial distance of 5 nm from the pore axis ($A_1$-$B_2$ and $A_1'$-$B_2'$), with its peak value increasing from ~0.69 to ~1 as $L_{cs}$ increases from 0 to 20 nm. In the absence of exterior surface charges (ICS case), a minimal flow rate is observed in the region of $B_2$-$A_2$ or $B_2'$-$A_2'$, due to the negligible fluid flow parallel to the uncharged exterior surface. However, when the exterior surface becomes charged, counterions on the entrance side enter the EDL region through $B_2$-$A_2$ and migrate into the nanopore along the exterior surface. On the exit side, beside entering the bulk from $A_1'$-$B_2'$, counterions can flow along the charged exterior surface and enter the bulk from $B_2'$-$A_2'$. The counterion inside EDLs near the membrane surface induces a parallel flow under the tangential electric fields. A local peak in the flow rate distribution appears at the position of $L_{cs}$, corresponding to the boundary of the charged exterior surface. These observations indicate that the charged exterior surface generates a tangential EOF component parallel to the membrane surface (Figures 4c and 4d), thereby enhancing the overall EOF rate through the nanopore.

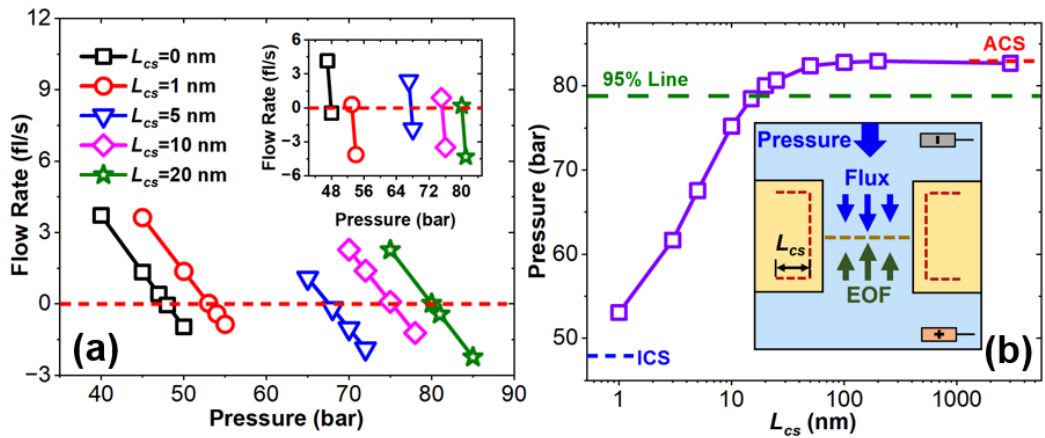

Figure 5. Effects of the width ($L_{cs}$) of the charged region on the output pressure induced by the electroosmotic flow through the nanopore. (a) Relationship between the output pressure and the flow rate at the nanopore cross-section. The inset zooms in on the region near zero flow rate. (b) Output pressure as a function of $L_{cs}$. The inset illustrates the nanofluidic simulation model, which integrates the coupled electric field and hydrostatic pressure.

In addition to the flow rate, the output pressure is another critical parameter for evaluating the performance of electroosmotic pumps,[22] as it represents the maximum hydraulic load the device can sustain during operation. In this study, we also investigated the impact of exterior surface charges on output pressure using a simulation model that couples the electric field and hydrostatic pressure. In the simulation, a hydrostatic pressure gradient opposing the electric field is applied to simulate an external load. The output pressure is defined as the pressure at which the flow rate through the central cross-section of the nanopore becomes zero.

From Figure 5a, at 1 V across the nanopore, the flow rate exhibits a linear correlation with the applied pressure. In the ICS case with neural exterior surfaces, a back pressure of 40 bar yields a forward flux of ~3.72 fl/s, whereas increasing the pressure to 48 bar reverses the flow to –0.05 fl/s. The output pressure of ~47.9 bar can be extracted from the intersection of the flow rate profile and the horizontal line at 0 flow rate. Following the same strategy, the output pressure for cases with various $L_{cs}$ can be obtained. Figure 5b illustrates the modulation of the output pressure by $L_{cs}$, following a trend similar to the variation of average EOF velocity with $L_{cs}$. As $L_{cs}$ increases from 0 to ~25 nm, the outlet pressure rises from ~47.9 to ~80.65 bar, by ~68%, which then approaches a saturation of ~82.9 bar. The enhancement in output pressure with increasing $L_{cs}$ can be attributed to the higher flow rate associated with larger $L_{cs}$ values. Expanding the charged region on the exterior surfaces significantly improves the output pressure, underscoring the critical role of exterior surface charges in enhancing the load-bearing capacity of electroosmotic pumps.

It is noteworthy that the normalized dependence of both the flow rate and output pressure on $L_{cs}$ exhibits nearly identical profiles, as shown in Figure S3. For the subsequent analysis, the effective charged widths are based on the variation in flow rate through the nanopores as a function of $L_{cs}$ under different conditions.

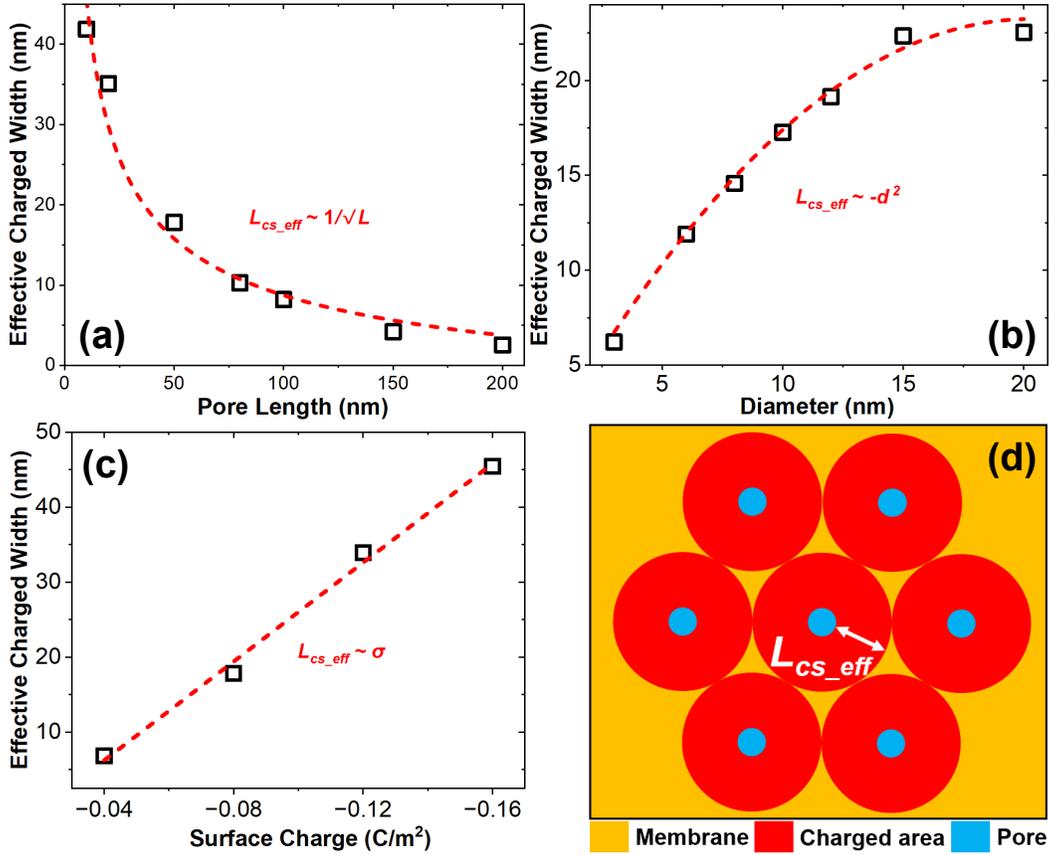

Figure 6. Dependence of the effective charged width ($L_{cs\_eff}$) on nanopore parameters, including (a) the pore length, (b) the pore diameter, and (c) the surface charge density. Default values were applied unless otherwise specified. (d) Schematic of the effective charge width ($L_{cs\_eff}$) on the exterior surface of the porous membrane. The separation between adjacent nanopores can be determined by twice of $L_{cs\_eff}$.

In practical applications, such as material transport and liquid propulsion,[1] porous membranes are typically considered the most feasible structure for achieving the desired flow rate. As the functional unit of porous membranes, the investigation of mass transport through individual nanopores can shed light on the optimal design of porous membranes. In the design of porous membranes, the separation between adjacent nanopores serves as a critical parameter that governs the transport characteristics of ions and fluid inside nanopores. Following the similar strategy in our previous research,[41, 42, 46] the effective width of charged exterior surfaces beyond the pore orifices bridges the study of fluid transport through individual nanopores and entire porous membranes. On porous membranes, if the separation between adjacent nanopores exceeds the effective width, the independent function of individual nanopores can be realized without any interference between nanopores, as shown in Figure 6d.[63, 64] Exploration of this effective width enables understanding of how exterior surface charges influence ion transport through nanopores, thereby providing guidance for the optimization of pore spacing.

Although a quantitative comparison between simulation results and existing experimental studies cannot be achieved, relevant nanofluidic research has demonstrated that charged exterior surfaces indeed play a significant role in regulating ion transport through nanopores.[44, 65, 66] When the pore length is reduced to the nanoscale (<100 nm), charged exterior surfaces can effectively promote the diffusion and migration of counterions within the pore.[16, 46, 67] The conductance of experimental verification can be extremely challenging under current technological and experimental conditions. Nilsson et al.[40] fabricated a 700-nm-long silicon nitride pore using focused ion beam (FIB) milling, achieved localized derivatization of pore entrances by controlling oxide ring growth, and preferentially functionalized the silicon oxide surface with silane chemistry. They observed that current intensity increases with the length of functionalized exterior surfaces. Recently, an experimental work conducted by Yazda et al. showed that on a 12-nm-thick hybrid hBN/SiN membrane, nanopores require a separation of ~500 nm to work independently during the osmotic energy conversion under a concentration gradient.[68] Their results are consistent with our previous simulations about the osmotic energy conversion with ultra-thin nanopores.[46] Based on our simulations, the effective charged width of ~250 nm aligns well with the optimal pore separation of ~500 nm in their report. So, the exploration of the effective charged width may provide a valuable parameter for the design of porous membranes.

Subsequently, the dependence of the effective charged width ($L_{cs\_eff}$) on the nanopore parameters and external conditions is explored systematically, including the length and diameter of nanopores, the surface charge density of the pore wall, the applied voltage, and the salt concentration. The values of $L_{cs\_eff}$ are determined as the corresponding $L_{cs}$ that delivers 95% of the flow rate observed in the ACS case under various conditions.

Figure 6a illustrates that the $L_{cs\_eff}$ declines nonlinearly as the pore length increases, which presents an inverse proportion to the square root of the nanopore length. As the length of the nanopore increases, the axial electric field strength weakens, reducing the driving force for fluid flow. Additionally, the flow resistance enhances in a longer nanopore. In this case, the influence of exterior surface charges on fluid flow becomes negligible when the nanopore length is sufficiently large. According to our results, for pore lengths exceeding 150 nm, the effect of exterior surface charges on EOF diminishes, and the $L_{cs\_eff}$ drops below 5 nm.

As shown in Figure 6b, $L_{cs\_eff}$ exhibits a parabolic relationship with the pore diameter, i.e., $L_{cs\_eff} \sim -d^2$. With the pore diameter enlarging from 3 to 15 nm, $L_{cs\_eff}$ increases from ~6.2 nm to ~22.5 nm. For 50-nm-long nanopores, as the diameter increases, the decreased confinement lowers the flow resistance across the nanopore, which enhances the EOF velocity. As a result, a wider charged exterior surface is needed to facilitate fluid transport. However, when the pore diameter exceeds 20 nm, the contribution of ion transport from the EDLs diminishes relative to the total ion transport through the nanopore, leading to a reduction in EOF. Consequently, $L_{cs\_eff}$ reaches its maximum value.

Surface charges are crucial for the appearance of EOF inside nanopores, which determine the characteristics of EDL near pore walls. Figure 6c plots the variation of $L_{cs\_eff}$ with the surface charge density. As the surface charge density increases from −0.04 to −0.16 C/m$^2$, $L_{cs\_eff}$ exhibits a linear growth from ~6.8 nm to ~45.5 nm, increasing by ~569%. This can be attributed to the enhanced EOF under a higher surface charge density, which necessitates a larger $L_{cs\_eff}$ to facilitate fluid transport through the nanopore.

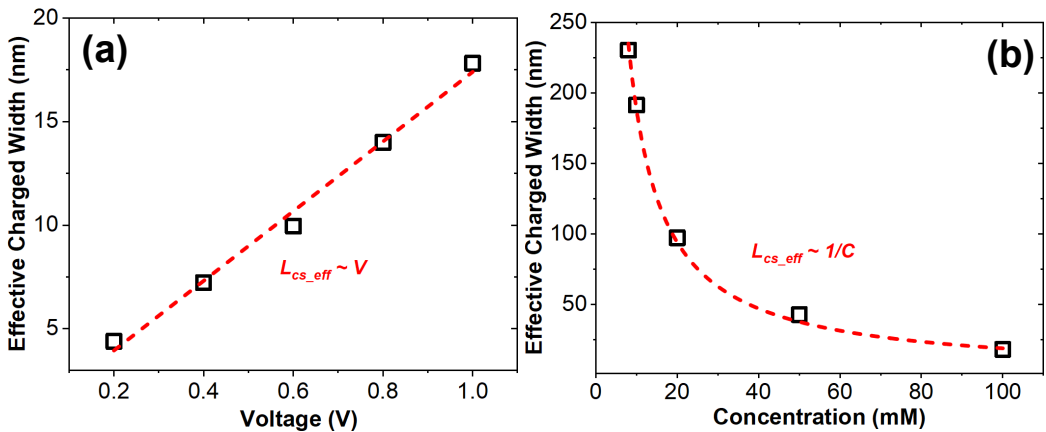

Figure 7. Dependence of the effective charged width ($L_{cs\_eff}$) on external conditions, including applied voltage (a) and salt concentration (b). Default values were applied unless otherwise specified.

EOF through nanopores is driven by the applied voltage, which also serves as the most convenient parameter to modulate the flow rate of EOF. From Figure 7a, $L_{cs\_eff}$ presents a linearly positive correlation with the applied voltage. As the applied voltage varies from 0.2 to 1 V, $L_{cs\_eff}$ increases from ~4 to ~18 nm, by ~350%. At higher voltages, enhanced EOF is induced by the stronger ion transport in EDL regions, which requires a larger $L_{cs}$ to accommodate the increased ion and fluid transport.

Electrolyte solutions are also crucial for microfluidic systems, as they supply both mobile ions and fluid for transport. For electrolyte solutions, the salt concentration is the key parameter that determines the Debye length.[7] Figure 7b presents the dependence of $L_{cs\_eff}$ on the salt concentration. In KCl solutions, as the concentration increases from 10 to 100 mM, $L_{cs\_eff}$ decreases nonlinearly from ~235 to ~20 nm, following $L_{cs\_eff} \sim 1/C$. This is mainly attributed to the diminished EOF due to the better screening of surface charges under higher salt concentration. Figure S4 shows the radial electric potential distribution at the center cross-section of the nanopore at different electrolyte concentrations. As the concentration increases, the absolute value of the surface potential decreases, which reduces the effective driving force for EOF and leads to decreased EOF velocity within the nanopore. As a consequence, a narrower charged exterior region is sufficient to maintain effective electroosmotic performance.

In this work, we focused on the modulation of EOF by exterior surface charges. Our results shown here are consistent with our previous studies about the modulation of ionic current through nanopores by exterior surface charges, where we considered different nanopore shapes, such as cylindrical[41] and conical nanopores[16], as well as different distributions of surface charges, including uniform[46] and bipolar charge distributions.[48] Based on our previous simulations, exterior surface charges can significantly enhance the ion transport through the nanopores under electric fields or concentration gradients. In each case, the $L_{cs\ eff}$ was obtained with different values. Also, the dependence of $L_{cs\ eff}$ on the parameter in different cases can have similar or different trends.[41, 46, 48]

Considering that the EOF flow measurement through single nanopores is extremely challenging, nanofluidic experiments with porous membranes may be attempted to be conducted in future studies. By systematically adjusting the pore spacing in porous membranes containing equal pore numbers and measuring the corresponding variations in the overall flow rate, the optimal width of the charged exterior surfaces can be determined at the maximum flow rate.

## IV. Conclusions

In this work, systematic simulations were conducted to investigate the modulation mechanism of EOF through cylindrical nanopores by exterior surface charges. For nanopores with a length of 50 nm and a diameter of 10 nm, the presence of exterior surface charges enhances both the counterion concentration and electric field strength within the nanopore, resulting in a ~55% increase in average EOF velocity and a ~73% increase in output pressure. By varying the charged ring area near both the pore entrance and exit, the effective width of the exterior charge surface is quantified as ~20 nm under default conditions. Then, the effective charge widths are explored under various nanopore parameters and applied conditions. Simulation results show that the effective area of the charged exterior surface near the pore orifice is directly proportional to the surface charge density and applied electric potential, inversely proportional to both pore length and solution concentration, and increases quadratically with pore diameter. These findings provide valuable theoretical insights and technical guidance for the development of high-performance membranes for practical EOF-based applications.

## SUPPLEMENTARY MATERIAL

See supplementary material for simulation details and additional simulation results.

## Acknowledgments

This research was supported by the Shandong Provincial Natural Science Foundation (ZR2024ME176), the Basic and Applied Basic Research Foundation of Guangdong Province (2025A1515010126), the Shenzhen Science and Technology Program (JCYJ20240813101159005), the National Natural Science Foundation of China (52105579), the Innovation Capability Enhancement Project of Technology-based Small and Medium-sized Enterprises of Shandong Province (2024TSGC0866), and the Key Research and Development Program of Yancheng City (BE2023010).

## AUTHOR DECLARATIONS

### Conflict of Interest

The author has no conflicts to disclose.

## DATA AVAILABILITY

The data that support the findings of this study are available from the corresponding author upon reasonable request.